\documentclass{article}

\usepackage{arxiv}

\usepackage[numbers]{natbib}
\usepackage[utf8]{inputenc} 
\usepackage[T1]{fontenc}    
\usepackage{hyperref}       
\usepackage{url}            
\usepackage{booktabs}       
\usepackage{amsfonts}       
\usepackage{nicefrac}       
\usepackage{microtype}      
\usepackage{lipsum}
\usepackage{graphicx}
\graphicspath{ {./images/} }

\usepackage{amssymb} 
\usepackage{amsmath}
\usepackage[ruled]{algorithm2e} 
\usepackage{multirow}
\usepackage{pgf}
\usepackage{arydshln} 
\usepackage{orcidlink}

\SetAlFnt{\small}
\SetAlCapFnt{\small}
\SetAlCapNameFnt{\small}
\SetAlCapHSkip{0pt}
\IncMargin{-\parindent}

\newcommand{\orcid}[1]{\href{https://orcid.org/#1}{\textcolor[HTML]{A6CE39}{\aiOrcid}}}

\title{What Drives Team Success? Large-Scale Evidence on the Role of the Team Player Effect}

\author{%
  \href{https://orcid.org/0000-0001-7679-8969}{\includegraphics[scale=0.06]{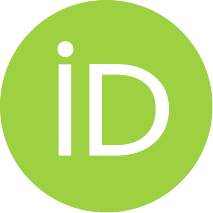}\hspace{1mm}Nico Elbert}\thanks{%
    For correspondence: \texttt{nico.elbert@uni-wuerzburg.de}}\\
  Julius-Maximilians-University Würzburg\\
  \texttt{nico.elbert@uni-wuerzburg.de}
  \And
  \href{https://orcid.org/0000-0003-3003-8475}{\includegraphics[scale=0.06]{orcid.pdf}\hspace{1mm}Alicia von Schenk}\\
  Julius-Maximilians-University Würzburg\\
  \texttt{alicia.vonschenk@uni-wuerzburg.de}
  \And
  \href{https://orcid.org/0000-0003-3458-3059}{\includegraphics[scale=0.06]{orcid.pdf}\hspace{1mm}Fabian Kosse}\\
  Julius-Maximilians-University Würzburg\\
  \texttt{fabian.kosse@uni-wuerzburg.de}
  \And
  \href{https://orcid.org/0000-0003-3384-8000}{\includegraphics[scale=0.06]{orcid.pdf}\hspace{1mm}Victor Klockmann}\\
  Julius-Maximilians-University Würzburg\\
  \texttt{victor.klockmann@uni-wuerzburg.de}
  \And
  \href{https://orcid.org/0000-0001-9847-3444}{\includegraphics[scale=0.06]{orcid.pdf}\hspace{1mm}Nikolai Stein}\\
  Julius-Maximilians-University Würzburg\\
  \texttt{nikolai.stein@uni-wuerzburg.de}
  \And
  \href{https://orcid.org/0000-0002-1761-9833}{\includegraphics[scale=0.06]{orcid.pdf}\hspace{1mm}Christoph Flath}\\
  Julius-Maximilians-University Würzburg\\
  \texttt{christoph.flath@uni-wuerzburg.de}
}
\begin{document}
\maketitle
\begin{abstract}
Effective teamwork is essential in structured, performance-driven environments, from professional organizations to high-stakes competitive settings. As tasks grow increasingly complex, achieving high performance requires not only technical proficiency but also strong interpersonal skills that enable individuals to coordinate effectively within teams. While prior research has identified social skills and familiarity as key factors in team performance, their combined effects—particularly in temporary teams—remain widely underexplored due to limitations in available data and experimental design. To address this gap, we analyze a large and detailed panel dataset of temporary teams operating in a highly competitive and structured environment. The dataset captures millions of interactions in the online real-time strategy game Age of Empires 2, where players are assigned to temporary teams through quasi-random matchmaking and must coordinate effectively under dynamic, high-pressure conditions. We isolate individual contributions to team success by comparing observed outcomes to predictions based on task proficiency. Our findings confirm the existence of a ‘team player effect,’ demonstrating that certain individuals systematically improve team outcomes beyond their technical skills alone. Moreover, we show that this effect is amplified by team familiarity—teams with prior shared experiences benefit more from the presence of team players. Additionally, social skills become increasingly important as team size grows, suggesting their role in overcoming coordination challenges in larger groups. These insights contribute to the literature on teamwork, social skills, and familiarity by demonstrating the robustness of the team player effect in a high-stakes, quasi-randomized setting. We further establish that social skills and familiarity interact in a complementary, rather than additive, manner. By highlighting how team familiarity strengthens the impact of socially skilled individuals and how coordination challenges scale with team size, our findings offer broader implications for team-based production in organizations and labor markets.
\end{abstract}

\begin{titlepage}

\vspace{1cm}
\setcounter{tocdepth}{1} 
\tableofcontents

\end{titlepage}


\section{Introduction}

Teamwork is a fundamental driver of success in structured, performance-driven environments, from professional organizations to high-stakes competitive settings. As tasks grow in complexity, achieving high performance requires not only technical proficiency but also strong interpersonal skills that enable individuals to coordinate, communicate, and collaborate effectively within teams.

In particular, temporary teams—such as those formed for specific projects or competitive endeavors—pose unique challenges. With limited time to establish coordination and trust, understanding the factors that drive team success becomes crucial \citep{bakker2013s,kim2023learning}.

Over the last few decades, jobs requiring social interaction have grown significantly as a share of the workforce \citep{deming2018skill}, while positions that emphasize technical skills alone have declined \citep{deming2017growing}. These shifts reflect broader changes in how work is organized, with firms placing greater emphasis on team-based production, where structured collaboration and problem-solving teams are leveraged to enhance productivity \citep{hamilton2003team}. Research suggests that beyond individual task proficiency, team-based production benefits from social and collaborative dynamics—such as knowledge-sharing, specialization, and adaptive problem-solving—especially in complex environments where incentive-based coordination is essential \citep{boning2007opportunity,guzzo1996teams,neuman1999team,ahearn2004leader}. Workers who excel at both technical and social tasks enjoy substantial labor market advantages, highlighting the complementarity between cognitive and social skills \citep{weinberger2014increasing}.

Teams often achieve greater efficiency than individuals working alone \citep{bloom2011human}. However, isolating the specific contribution of social skills to team performance remains a significant challenge. Studies have shown that certain individuals consistently improve team performance beyond what would be expected based on their technical skills alone. These “team players” exemplify how social skills enhance team coordination and productivity \citep{weidmann2021team}, a phenomenon referred to as the team player effect. Further, research suggests that factors such as specialization, social familiarity among team members, and diversity in skill, knowledge, and experience are critical drivers of effective collaboration and team outcomes \citep{ching2021extemporaneous,ching2024competitive,hollenbeck1995multilevel,zhang2023know,harrison2003time}.
 \\

Temporary teams play an increasingly vital role in modern organizations, and the effective coordination of interdependent actors is a core challenge of organizational design \citep{becker1992division}. However, our understanding of how individual and social skills interact within these teams remains limited. While the team player effect and other key factors have been analyzed in isolation, their interplay is widely underexplored, largely due to the lack of accessible large-scale data or realizable experimental settings. Studies conducted in laboratory environments often focus on simplified tasks that fail to capture the complexity and interdependencies inherent in real-world team dynamics \citep{salas2008teams}.

To address this gap, we analyze a large and detailed panel dataset of temporary teams operating in a highly competitive and structured environment. Our setting provides a millionfold execution of an experiment in which team members are exogenously assigned to teams through quasi-random matchmaking \citep{goetz2022peer,deng2021globally}. The dataset captures millions of interactions in the online real-time strategy game Age of Empires 2 (AoE2), where players must coordinate effectively under dynamic, high-pressure conditions. Like real sports, which are widely used in organizational research \citep{fonti2023using,massaro2020intellectual,mach2010differential}, e-sports or competitive online games offer structured settings and granular data that allow for detailed analysis of teamwork and performance. They provide a naturalistic environment \citep{chen2023platform} where intrinsically motivated players exhibit goal-oriented decision-making, develop adaptive strategies, and engage in long-term skill learning \citep{liu2013digital, huang2019level}. These games capture complex organizational behaviors, with players collaborating in dynamic teams, taking on specialized roles, and navigating shared goals under pressure \citep{petter2020gaming, freeman2019understanding}.
Additionally, they provide opportunities to assess individual abilities, track performance metrics, and analyze decision-making preferences \citep{kunn2022cognitive, dertwinkel2024skewness,allen2024using}. As a result, researchers have increasingly recognized the potential of e-sports and competitive online games as a laboratory for testing organizational theories, given its parallels with real-world settings \citep{waguespack2018cultural, clement2023missing, coates2020managers}. \\

In this paper, we develop and test an empirical method that integrates the approaches of \citet{weidmann2021team} and \citet{ching2021extemporaneous} to identify individual contributions of individual task proficiency and social skills to team performance. We further examine whether the influence of the team player effect depends on team size and team familiarity, as structured coordination may become more relevant in larger or more familiar teams.

We first assess individual abilities by analyzing competitive solo interactions, where players complete the same tasks independently. This approach allows us to isolate metrics for mechanical skill and ultimately task proficiency in a controlled setting without the influence of team dynamics.

Next, we evaluate team performance in a competitive environment where individuals are quasi-randomly assigned to temporary teams. By incorporating individual performance metrics and contextual factors—such as a player’s functional familiarity, defined as their prior experience within the specific setting—we account for all individual fixed effects related to task proficiency that influence team outcomes.

We then identify instances where specific players are consistently part of teams that exceed predicted outcomes. Following the approach of \citep{weidmann2021team}, we attribute this improvement—unexplained by individual skills or functional familiarity—to their unique ability to enhance team coordination and productivity. This phenomenon is defined as the team player effect, and we refer to these individuals as team players.

On an independent set of observations, we incorporate the team player effect and the social familiarity metric introduced by \citet{ching2021extemporaneous}—which captures the extent of prior shared experiences between teammates—to evaluate the explanatory power of each feature and analyze how their interplay contributes to team performance.

Our findings confirm the existence of the team player effect in a high-stakes, dynamic environment, aligning with prior research that identifies individuals who systematically enhance team performance beyond their task proficiency alone \citep{weidmann2021team}. We show that the impact of socially skilled individuals is not only significant but also moderated by team familiarity—teams with prior shared experiences benefit more from the presence of team players, suggesting a complementary relationship between social skills and familiarity. Furthermore, we find that the role of social skills intensifies as team size increases, indicating that team players become increasingly important in larger teams where coordination challenges are more pronounced.

These results extend previous research by demonstrating the robustness of the team player effect in a large-scale, competitive setting, moving beyond controlled laboratory environments to quasi-randomly assigned team interactions. Our findings also highlight the interplay between social skills and team familiarity, showing that their effects are not merely additive but complementary, with the strongest performance gains observed in familiar teams. Lastly, we provide new evidence that team size moderates the impact of social skills, underscoring their role in scaling coordination in larger groups.

Together, these insights contribute to a broader understanding of how team familiarity, specialization, and social skills interact in dynamic team environments, offering valuable implications for team-based production in organizations.

The remainder of the paper proceeds as follows. Section \ref{sec:related_work} reviews the relevant literature. Section \ref{sec:institutional_context} introduces the study context. Section \ref{sec:data} describes the sourcing of our dataset and the construction of key variables. Section \ref{sec:study_design} details our methodology. Section \ref{sec:results} presents our findings. Section \ref{sec:discussion} interprets the results, and Section \ref{sec:conclusion} concludes.  

\section{Related Work} \label{sec:related_work}

A fundamental challenge in team-based production is how individuals coordinate effectively when placed into temporary teams with little to no prior interaction. Unlike stable teams, temporary teams must rapidly establish coordination mechanisms without prolonged periods of adjustment \citep{bakker2013s, kim2023learning,akcsin2021learning}. While previous research in organizational settings highlights that familiarity and trust among teammates can enhance team efficiency \citep{hollenbeck1995multilevel,massaro2020intellectual,mach2010differential,harrison2003time,maynard2019really}, teams often cannot rely on familiarity alone, making it crucial to understand other factors that drive performance.

One key determinant of team success in unfamiliar settings is social skill \citep{morgeson2005selecting,neuman1999team}. \citet{weidmann2021team} introduce the \textit{team player effect}, demonstrating that certain individuals systematically improve team performance beyond what their technical skills alone would predict. Their study, conducted in a controlled laboratory setting, isolates social skills as a key determinant of team performance, showing that team players enhance group coordination and effort allocation.

To measure the team player effect, \citet{weidmann2021team} designed an experiment where participants were first assessed on their individual ability to complete various problem-solving tasks, including memory, optimization, and spatial reasoning. Participants were then randomly assigned to multiple teams of three, ensuring that they worked with different teammates across multiple sessions. This repeated randomization allowed the researchers to estimate individual contributions to team performance while controlling for task-specific skills.

The expected team performance was predicted based on the sum of the individual skill levels of its members. The team player effect was identified as the extent to which a team's observed performance exceeded this predicted value when a specific individual was part of the group. The residual differences, aggregated across multiple team assignments, provided an estimate of each individual's team player index. Their findings showed that social intelligence—measured through the \textit{Reading the Mind in the Eyes Test (RMET)} \citep{baron2001reading}—was a strong predictor of team player status, whereas traditional cognitive measures such as IQ provided no additional explanatory power. 

These results highlight that teamwork efficiency depends not only on functional expertise or task proficiency but also on interpersonal skills that facilitate structured collaboration. By leveraging randomization and predictive modeling, \citet{weidmann2021team} provided robust evidence that social skills contribute meaningfully to team outcomes, supporting the argument that non-cognitive skills are crucial in collaborative settings.

While individual social skills improve team coordination, prior shared experiences and role-specific expertise—collectively referred to as \textit{familiarity}—also play a crucial role in shaping performance outcomes. \citet{ching2021extemporaneous, ching2024competitive} distinguish between two key dimensions of familiarity: \textit{social team familiarity} (henceforth \textit{team familiarity}) and \textit{functional familiarity}.

Team familiarity captures repeated interactions between specific teammates, fostering implicit coordination and reducing communication costs through shared experiences. In contrast, functional familiarity arises from individuals repeatedly performing in similar roles or contexts, regardless of their specific teammates, leading to deeper task proficiency and an improved understanding of role interdependencies. \citet{ching2021extemporaneous} find that teams benefit most when team familiarity is combined with structured role specialization, as this facilitates adaptive decision-making and minimizes coordination overhead. Their research highlights the complementarity between social and functional familiarity, showing that coordination in dynamic environments depends not only on interpersonal experience but also on the stability of role expectations.

While \citet{weidmann2021team} identify the team player effect through a controlled laboratory experiment with randomized team assignments, \citet{ching2021extemporaneous, ching2024competitive} adopt an empirical approach within the competitive online game \textit{DOTA2}. Such empirical studies of real-world competitive environments require domains that naturally feature structured team interactions, goal-oriented behavior, and measurable performance outcomes—characteristics inherent to sports \citep{fonti2023using,massaro2020intellectual,mach2010differential} and competitive gaming or esports \citep{reitman2020esports,coates2020managers,kou2014playing}.

Competitive gaming and esports are closely related but not interchangeable terms. Competitive gaming broadly encompasses all structured digital competitions, ranging from digital adaptations of traditional games like chess and poker to highly complex multiplayer games with ranking systems and performance-based incentives. Esports, by contrast, refers specifically to the professionalized tier of competitive gaming, characterized by structured teams, coaching, sponsorships, and financial stakes with prize pools reaching up to 40 million USD\footnote{``https://www.esportsearnings.com/tournaments’’ on February 4th, 2025}. Despite this distinction, both competitive gaming and esports share the same digital game environments, which inherently generate vast amounts of structured data, including real-time player actions, ranking systems, and social interactions, which enable precise empirical analysis of individual and team performance on every level of play. Previously, such detailed data was primarily available in professional sports \citep{abeza2023big} or in environments with discrete and well-defined state spaces, such as chess \citep{kunn2023indoor}.

A defining characteristic of competitive gaming is its structured incentive system, which fosters rational and goal-directed behavior among players \citep{sepehr2018understanding,yee2006motivations}. Players engage in competitive matches with clear objectives, where success is quantified through performance metrics such as rankings, win rates, and skill ratings. These ranking systems create strong intrinsic and extrinsic incentives, motivating players to refine their skills and optimize decision-making \citep{liu2013digital, bruhlmann2020motivational}. Competitive motivation is often situational and dispositional, meaning players actively seek out challenging environments and systematically develop expertise over time \citep{yee2016gamer,williams2008plays}. This distinguishes esports from more traditional experimental settings, where participants may lack intrinsic motivation to perform optimally.

The availability of detailed, structured data in competitive gaming environments has enabled researchers to analyze a wide range of behavioral and strategic dynamics, such as adaptation to rule changes, engagement responses to platform adjustments, and decision-making under varying competitive conditions \citep{chen2023platform, clement2023missing}.

Unlike laboratory experiments that rely on artificially constructed teams, competitive gaming provides naturally occurring interactions where players engage with both familiar teammates and quasi-randomly assigned strangers. Research leveraging these settings has shown that peer effects influence in-game decision-making, with interactions among friends and strangers shaping behavioral responses \citep{deng2021globally}.

Overall, competitive gaming provides a highly structured, incentive-driven environment that mirrors team-based work in traditional organizations. The combination of dynamic interactions, quantifiable performance metrics, and varying degrees of team familiarity makes competitive gaming a valuable domain for advancing research on teamwork, social skills, and coordination.


\section{Institutional Context: Age of Empires 2} \label{sec:institutional_context}

We analyze team performance in the context of AoE2, a real-time strategy game that combines high levels of individual decision-making, team coordination, and dynamic adaptation. AoE2 is particularly well-suited for our analysis due to its high task complexity and the presence of temporary teams. Matches are structured as zero-sum competitions where players must simultaneously manage multiple dimensions of gameplay, including resource acquisition, military engagements, and strategic decision-making, under significant time constraints. \citep{bayrak2021strategic}

The interdependence between team members, combined with the quasi-randomized nature of team assignments \citep{deng2021globally} in competitive matches, allows us to examine the impact of individual task proficiency, social skills and team familiarity under controlled conditions.

The game's design also enables precise measurement of key constructs relevant to this study. Each player is uniquely identifiable across matches, allowing us to track individual histories and construct detailed metrics for mechanical skill (e.g., Actions Per Minute, eAPM), tactical skill (e.g., Solo Elo), team familiarity, and functional familiarity. These metrics, combined with extensive metadata on matches, roles, and positions, form the foundation of our empirical analysis.

\subsection{Match Types}


AoE2 matches are categorized as either ranked or unranked, with each format available in solo (1v1) and team (2v2, 3v3, 4v4) modes. Ranked matches are distinguished by their competitive nature, as players’ performance directly contributes to their Elo rating—a central measure of skill that reflects their track record over time. Separate Elo ratings are maintained for solo and team matches, allowing for distinct evaluations of individual and collaborative performance. The Elo system not only ranks players but also serves as a probabilistic estimator of match outcomes, with rating differentials used to infer expected win probabilities and ensure balanced matchmaking. This system not only quantifies individual success but also serves as a key motivator, driving players to improve their skills \citep{sepehr2018understanding, bruhlmann2020motivational}. Unranked matches, by contrast, offer a more casual setting for experimentation and skill refinement without the risk of damaging the Elo ranking. Our analysis focuses exclusively on ranked matches. This ensures that the behaviors and outcomes observed reflect the serious intent with which players approach these competitive settings. In addition to the modes, matches are further differentiated by game settings, including:

\paragraph{Positions and Roles.} In team matches, players are exogenously assigned starting positions on the map, which implicitly determine their roles within the team. These roles reflect a natural division of labor based on spatial positioning and task specialization. Players positioned closer to opponents—typically referred to as flanks—initially operate in relatively isolated conditions, requiring independent decision-making and rapid adaptability to opponent actions. In contrast, centrally positioned players—often referred to as pockets—fulfill more interdependent and coordinative roles, focusing on structural development and providing support to enable the flanks’ offensive or defensive strategies. Their actions are closely tied to team objectives, requiring broader situational awareness of both teammates’ and opponents’ positions.



\paragraph{Civilizations.} Players select from a wide array of civilizations, each with unique attributes. These differences drive strategic diversity, as players adjust their gameplay to exploit their civilization’s strengths while mitigating its weaknesses. In team matches, civilization selection is often aligned with a player’s assigned role within the team, enhancing coordination by balancing offensive, defensive, and resource-focused strategies.

\paragraph{Maps.} Matches take place on procedurally generated maps that vary significantly in terrain, resource distribution, and strategic chokepoints. These variations introduce additional layers of complexity, requiring players to adapt their strategies to the spatial arrangement of resources, their proximity to opponents, and the defensive or offensive opportunities provided by the map layout.

\subsection{Team Formation and Randomization}

Ranked matches in AoE2 are organized through a matchmaking algorithm that assigns players to temporary teams based on their Teanm Elo ratings. This system ensures competitive balance by minimizing skill disparities between opposing teams, creating an environment suitable for analyzing team performance. Teams can consist of entirely random players or include premade groups of players who join together. 

These premade teams may form from prior in-game interactions or preexisting social connections outside the game.  Over multiple engagements, players may build familiarity and form social bonds, enhancing coordination and reinforcing team performance \citep{ching2021extemporaneous, ching2024competitive, freeman2019understanding}. A team may consist of a mix of premade groups and randomly assigned players, reflecting the diverse compositions possible within the matchmaking system.

The matchmaking algorithm does not explicitly account for whether teams are premade or entirely random but balances overall skill by pairing teams with comparable Elo ratings. This ensures fairness across matches while introducing exogeneity into teammate assignments, providing natural variation in team compositions. By controlling for team familiarity among team members, we can distinguish matches with completely random teams from those involving partially or fully premade teams. This control enables us to isolate the effects of the team player effect and team familiarity on match outcomes.

Matchmaking also involves trade-offs between competitive balance and waiting times. To minimize delays, the system restricts the matchmaking pool based on player availability, which is further influenced by time zones. Players from similar regions are prioritized to reduce latency. These constraints, while practical, impose limits on the diversity of potential opponents and teammates, influencing the composition of the matchmaking pool.

By introducing variation through modes, roles, civilizations, and maps, alongside the described randomization, AoE2 provides a rich and dynamic environment for studying team dynamics. The combination of exogenously assigned teammates, diverse gameplay settings, and measurable constructs such as familiarity and coordination allows for robust analysis of how temporary teams adapt and perform under varying conditions. This setting enables the identification of causal relationships that are otherwise challenging to isolate in observational data.
\section{Data and Variable Construction}
\label{sec:data}

Our study analyzes ranked solo and team-based matches in AoE2, comprising 3,785,939 matches, of which 1,623,828 are team-based. The dataset captures detailed information on player performance, team composition, and functional familiarity, enabling us to examine how team familiarity and the team player effect influence team performance.

\subsection{Sourcing}
The dataset was sourced from the analytics platform \textit{aoe2insights.com} and centers on a core set of 14,000 focal players who participated in at least 100 ranked solo and 100 ranked team matches between November 2019 and December 2023. In total, these players contributed to 1,623,828 team matches across 1,788,095 unique team formations, with the broader dataset encompassing 137,412 individual players.

These players form temporary teams, represented as \(g \subseteq P\), where \(P\) is the set of all players. The set of all teams \(G \subseteq 2^P\) consists of teams \(g\), which participate in matches \(m \in M\) at specific timestamps \(t \in T\). Formally, the dataset is defined as:
\[
M = \{m_1, m_2, \dots, m_n \}, \quad G = \{g_1, g_2, \dots, g_k \}, \quad P = \{p_1, p_2, \dots, p_l \}
\]
where matches \(m\) involve teams \(g \subseteq P\) and occur at timestamps \(t \in T\).

The full dataset includes metadata on 1,623,828 team matches, capturing a comprehensive view of interactions across solo and team modes. The collected data includes:
\begin{itemize}
    \item \textbf{Match Metadata:} Game mode (2v2, 3v3, 4v4), map type, civilizations used, and match outcome.
    \item \textbf{Team Composition:} Team members \(p \in g\) and their associations within each match.
    \item \textbf{Player Metrics:}
    \begin{itemize}
        \item Historical Solo Elo (\(SElo_t\)): Rating for individual tactical skill.
        \item Historical Team Elo (\(TElo_t\)): Rating reflecting both individual and team performance.
        \item Sampled game recordings for eAPM and other mechanical metrics.
        \item Player ID and country of origin.
    \end{itemize}
\end{itemize}

\begin{figure}[!h]
    \centering
    \begin{minipage}{0.48\textwidth}
        \includegraphics[width=\textwidth]{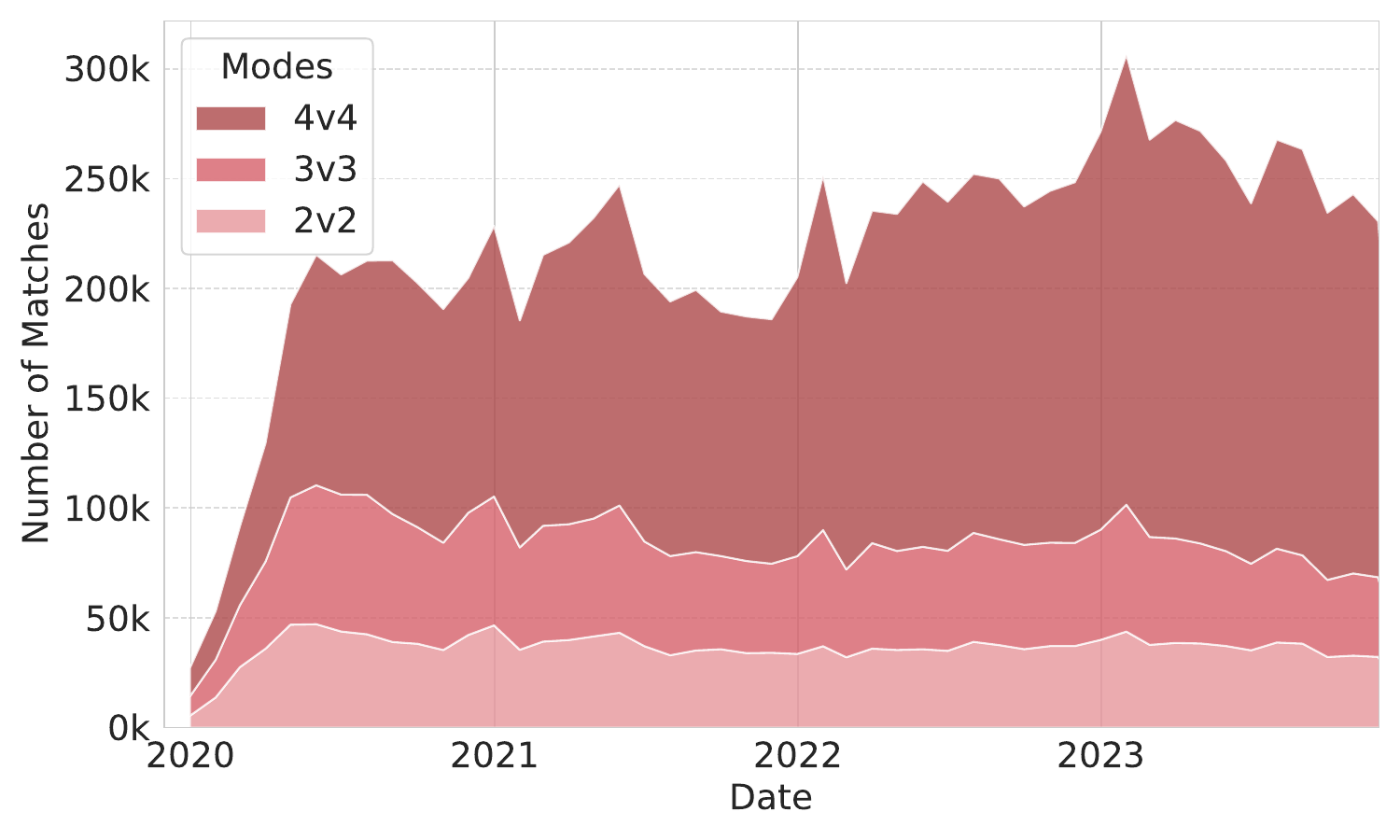}
    \end{minipage}
    \hfill
    \begin{minipage}{0.48\textwidth}
        \includegraphics[width=\textwidth]{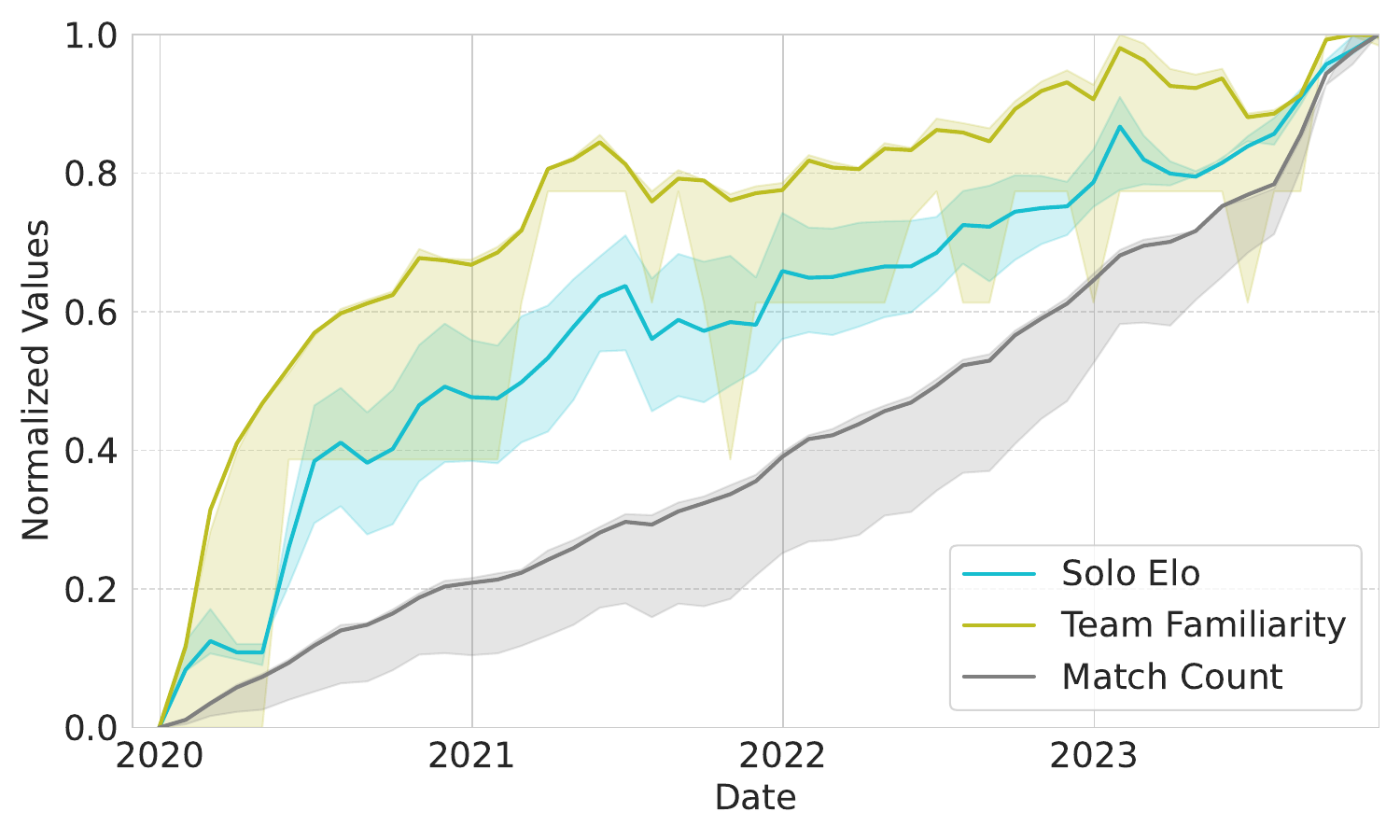}
    \end{minipage}
    \caption{(Left) Distribution of Team Matches per Mode over Time (monthly); (Right) Standardized Mean Feature Values with Quantiles within Collected Matches over Time (monthly).}
    \label{fig:timelines}
\end{figure}


\subsection{Variables and Measures}

In this study, we utilize four key variables to quantify individual and team-level contributions to performance: mechanical skill in actions per minute (eAPM), tactical skill (Solo Elo), functional familiarity, and team familiarity. For analysis, team-level features are aggregated, and differences between teams are calculated as deltas, facilitating a detailed comparison of team performance. These variables are formalized as follows:

\subsubsection{Mechanical Skill (eAPM)}

Mechanical skill is captured using eAPM, which measures the number of effective game actions performed by a player per minute. eAPM is extracted from detailed game recordings and provides a proxy for a player’s efficiency in executing tasks. Formally, for a player \(p \in P\) in match \(m \in M\), eAPM is defined as:
\[
eAPM_{p,m} = \frac{\text{Total Effective Actions Performed by } p}{\text{Total Duration of Match } m}
\]
This variable isolates the speed and execution aspect of individual performance.

At the team level, eAPM is averaged across all players \(p \in g_m\) within a team \(g_m \subseteq P\) for match \(m\):
\[
eAPM_t(g_m) = \frac{1}{|g_m|} \sum_{p \in g_m} eAPM_{p,m},
\]
capturing the average mechanical skill of the team members.

\subsubsection{Tactical Skill (Solo Elo)}

Solo Elo (\(SElo\)) measures a player’s tactical skill in 1v1 matches. It reflects the player’s ability to adapt strategies dynamically and make effective decisions under competitive conditions. Unlike team Elo, Solo Elo captures individual skill without the confounding influence of team coordination. The historical Solo Elo \(SElo_{p,t}\) for a player \(p\) at time \(t\) is included as a baseline feature to control for tactical ability:
\[
SElo_{p,t} \in \mathbb{R}, \quad \text{where higher values indicate greater tactical skill.}
\]

At the team level, Solo Elo is averaged across all players \(p \in g_m\) within a team \(g_m \subseteq P\) for match \(m\) at time \(t\):
\[
SElo_t(g_m) = \frac{1}{|g_m|} \sum_{p \in g_m} SElo_{p,t},
\]
capturing the average tactical skill of the team.
\subsubsection{Functional Familiarity}

Functional familiarity measures a player’s accumulated experience within specific contexts, such as game modes, maps, or civilizations. This variable represents a form of specialized human capital that enables players to efficiently perform tasks within familiar environments. For a player \(p\) and context \(c \in C\), functional familiarity up to time \(t\) is computed as:
\[
FFam_{p,c,t} = \log \left( 1 + \sum_{m \in M_c^t} 1 \right),
\]
where \(M_c^t\) is the set of matches in context \(c\) played by \(p\) up to time \(t\). This log transformation ensures that diminishing returns are applied to extensive experience, preventing overemphasis on highly specialized players.

At the team level, functional familiarity is averaged across all players \(p \in g_m\) within a team \(g_m \subseteq P\) for match \(m\) at time \(t\):
\[
FFam_t(g_m) = \frac{1}{|g_m|} \sum_{p \in g_m} FFam_{p,c,t},
\]
capturing the average contextual experience of all team members.


\subsubsection{Team Familiarity}


Team Familiarity captures the extent of prior interactions between teammates. It is calculated as the log-transformed mean of pairwise familiarities among all team members, ensuring that the measure remains comparable across teams of different sizes while limiting the marginal utility in highly familiar teams. For a team $g_m \subseteq P$ in match $m$ at time $t$, team familiarity is defined as:

\[
TFam_t(g_m) = \log \left( 1 + \frac{1}{|g_m|(|g_m|-1)} \sum_{\substack{p_1, p_2 \in g_m \\ p_1 \neq p_2}} Fam_t(p_1, p_2) \right),
\]

where $Fam_t(p_1, p_2)$ represents the total number of matches where players $p_1$ and $p_2$ appeared on the same team up to time $t$. This measure captures the degree to which team members have developed implicit coordination through repeated interactions, while accounting for differences in team size.

\subsection{Descriptive Statistics}

The dataset is characterized by a high prevalence of temporary teams, with the majority forming only once. The median number of appearances per team is 1, with the 75th percentile also at 1, indicating that most teams are short-lived. Only at the 95th percentile do teams appear at least three times, and the 99th percentile reaches 10 appearances, highlighting the rarity of persistent team compositions. Individual players participate more frequently, with a median of 6 matches and a highly skewed distribution—while 75\% of players have at most 31 appearances, the 95th percentile reaches 324, and the 99th percentile surpasses 1,100.

\begin{table}[h!]
\smaller
    \caption{Pearson Correlation Coefficients Between Features}
    \label{tab:correlation_table}
    \begin{minipage}{\columnwidth}
        \begin{center}
            \begin{tabular}{l @{\hskip 10pt} l @{\hskip 10pt} c @{\hskip 10pt} c @{\hskip 10pt} c @{\hskip 10pt} c @{\hskip 10pt} c @{\hskip 10pt} c @{\hskip 10pt} c}
                \toprule
                & \textbf{Feature} & \textbf{eAPM} & \textbf{SElo} & \textbf{Matches} & \textbf{Map} & \textbf{Civ.} & \textbf{Team Fam.} & \textbf{Win} \\
                \midrule
                \multirow{2}{*}{\textit{Solo Features}} 
                & eAPM & 1.000 & 0.266 & 0.062 & 0.087 & 0.072 & 0.175 & 0.148 \\
                & SElo & 0.266 & 1.000 & 0.183 & 0.124 & 0.100 & 0.084 & 0.291 \\
                \hdashline
                \multirow{3}{*}{\textit{Functional Familiarity}} 
                & Matches & 0.062 & 0.183 & 1.000 & 0.642 & 0.583 & 0.307 & 0.088 \\
                & Map & 0.087 & 0.124 & 0.642 & 1.000 & 0.688 & 0.368 & 0.110 \\
                & Civilization & 0.072 & 0.100 & 0.583 & 0.688 & 1.000 & 0.307 & 0.087 \\
                \hdashline
                \multirow{2}{*}{} 
                & Team Familiarity & 0.175 & 0.084 & 0.307 & 0.368 & 0.307 & 1.000 & 0.063 \\
                & Win & 0.148 & 0.291 & 0.088 & 0.110 & 0.087 & 0.063 & 1.000 \\
                \bottomrule
            \end{tabular}
        \\
        \vspace{5pt}
        \footnotesize \emph{Note:} All p-values $\leq$ 0.01. 
        \end{center}
    \end{minipage}

\end{table}

Figure \ref{fig:timelines} provides insights into the temporal evolution of match activity and tracks the standardized progression of key performance-related attributes over time. Table \ref{tab:correlation_table} presents Pearson correlation coefficients between key features. Mechanical skill (eAPM) and tactical skill (Solo Elo) show a moderate correlation (0.266), while functional familiarity variables exhibit the highest internal correlations. Win rate is most strongly associated with Solo Elo (0.291) and eAPM (0.148), highlighting the role of both tactical and mechanical skills, though the modest correlations suggest that coordination and team dynamics are also critical. The Variance Inflation Factor (VIF) is $<$ 5 for all features, except Functional Familiarity, which has the highest internal correlation.

\begin{table}[h]
\smaller
    \caption{Mean Team APM and Correlation with Win Probability Across Elo Bins}
    \label{tab:elo_apm_table}
    \begin{minipage}{\columnwidth}
        \begin{center}
            \begin{tabular}{lccc}
                \toprule
                \textbf{Elo Bin} & \textbf{Mean Team APM} & \textbf{Corr. ($\Delta$APM–Win)} \\
                \midrule
                $<$1000 & 31.20 & 0.132 \\
                1000–1500 & 38.14 & 0.132 \\
                1501–2000 & 44.30 & 0.156 \\
                $>$2000 & 50.79 & 0.185 \\
                \bottomrule
            \end{tabular}
        \\
        \vspace{5pt}
        \footnotesize \emph{Note:} All p-values $\leq$ 0.01. 
        \end{center}
    \end{minipage}
\end{table}

In addition to overall correlations, we observe systematic variation in mechanical skill and its association with match outcomes across Elo tiers. As shown in Table~\ref{tab:elo_apm_table}, mean team eAPM increases consistently with tactical rating, from 31.2 among teams below 1000 Elo to 50.8 among those above 2000. The correlation between within-match eAPM differences and winning remains stable at 0.13 in lower tiers but rises to 0.18 in the highest bin, suggesting that mechanical execution becomes increasingly decisive at higher skill levels. 








\section{Study Design}
\label{sec:study_design}

\begin{figure}[!h]
	\includegraphics[width=\textwidth]{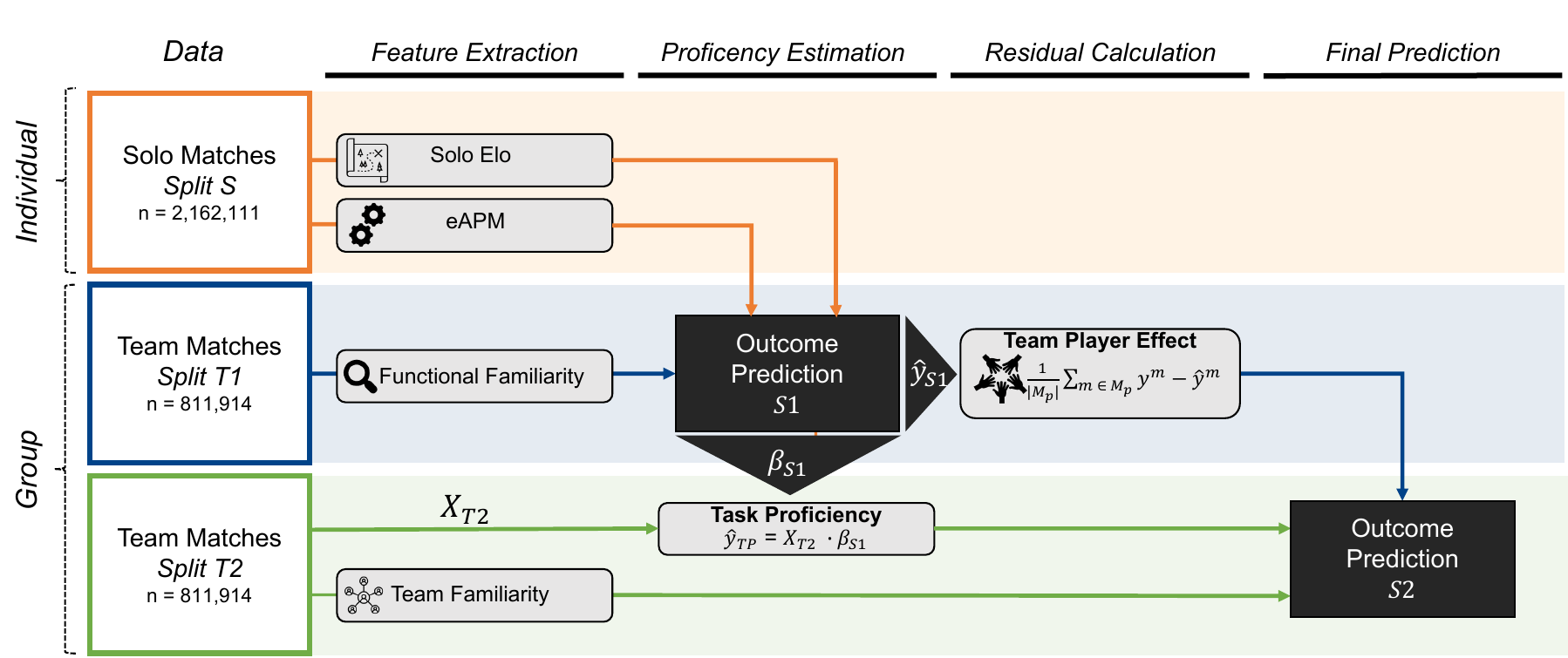}
	\caption{Pipeline for Feature Extraction, Residual Calculation, and Final Prediction}
	\label{fig:pipeline}
\end{figure}

Our study is structured around three distinct data splits: individual interactions (Split S) and group interactions (Splits T1 and T2), as illustrated in Figure \ref{fig:pipeline}. These splits are utilized in a sequential framework consisting of feature extraction, task proficiency estimation, residual calculation, and final outcome prediction.

To assess the contribution of each feature, we estimate a sequence of independent logistic regressions, each progressively incorporating additional explanatory variables relevant to team performance. This stepwise approach allows us to isolate the effects of individual features while ensuring comparability across models. In the first stage (S1), we evaluate the individual contributions of features to task proficiency and derive the team player effect. In the second stage (S2), we incorporate these components alongside team familiarity to assess their joint influence on team performance and their interactions.

\paragraph{\textbf{Split S:} Individual Features.} The first split consists exclusively of solo matches (\(n = 2,162,111\)) and serves to extract the individual skill features Solo Elo and eAPM. These features capture mechanical and tactical proficiency at the individual level, independent of team interactions, providing a baseline for evaluating task proficiency.

\paragraph{\textbf{Split T1:} Estimating Task Proficiency and Team Player Effect.} 
The second split, T1 (\(n = 811,914\)), is used to:
\begin{enumerate}
    \item Calculate Functional Familiarity as a measure of players’ prior exposure to specific game settings, capturing experience-driven advantages.
    \item Fit a logistic regression model to predict match outcomes using individual skill features from Split S (Solo Elo and eAPM) and Functional Familiarity. The estimated coefficients from this model define the composition of Task Proficiency, represented by the predicted outcome \(\hat{y}_{S1}\).
    \item Compute the Team Player Effect as the residual \( r = y - \hat{y}_{S1} \), capturing the portion of team performance unexplained by task proficiency.
\end{enumerate}

\paragraph{\textbf{Split T2:} Feature Application and Final Prediction.} 
The third split, T2 (\(n = 811,914\)), is used to:
\begin{enumerate}
    \item Construct the feature vector \(X_{T2}\) incorporating team familiarity and the estimated features from Split T1 (task proficiency and team player effect).
    \item Apply the estimated coefficients (\(\beta_{S1}\)) from Split T1 to the corresponding feature values in Split T2 to compute task proficiency for teams in Split T2.
    \item Fit the final outcome prediction model using task proficiency, team familiarity, and the team player effect, enabling an analysis of how these factors jointly influence team performance.
\end{enumerate}


\subsection{Win Prediction}

Our win predictions are based on the normalized delta of team features listed in Table \ref{tab:correlation_table}, representing the difference between the values of each feature for the two competing teams in a given match. By focusing on the feature deltas, we aim to capture the relative advantage of one team over the other across various dimensions such as individual skills, familiarity, and team dynamics.

To evaluate the predictive performance of our models, we apply an 80:20 train-test split on the dataset. This ensures that the model is trained on 80\% of the data and evaluated on the remaining 20\%, providing an unbiased measure of its generalization capability. The training process involves fitting a logistic regression model on the train set using the specified feature sets as outlined in Figure \ref{fig:pipeline}. Logistic regression is particularly suited for this task, as it directly models the probability of a binary outcome, which in this case is whether a team wins (\(y = 1\)) or loses (\(y = 0\)). To account for the fact that teams can appear multiple times in the dataset, we cluster standard errors at the team level, ensuring that our inference accounts for within-team dependencies.

The trained models are then evaluated on the test set using accuracy as the primary performance metric. Accuracy is calculated as the proportion of correctly predicted match outcomes out of the total matches in the test set. This straightforward yet effective metric allows for direct comparisons of predictive performance across different feature specifications.

To estimate the probability of team \( g \) winning a given match, we model the outcome as a function of feature differences.

\begin{equation}
P(y_m = 1 \mid \Delta \mathbf{X}_m) = \sigma\left( \beta_0 + \sum_{j} \beta_j \Delta X_{mj} + \varepsilon_m \right),
\end{equation}

where:
\begin{itemize}
    \item $y_m$ is the binary match outcome for match $m$, defined from the perspective of team $g_A$ ($y_m = 1$ if $g_A$ wins, $y_m = 0$ otherwise),
    \item $\sigma(\cdot)$ represents the logistic (sigmoid) function,
    \item $\Delta X_{mj}$ denotes the difference in feature $j$ between the two competing teams, defined as $\Delta X_{mj} = X_{mj}^{(g_A)} - X_{mj}^{(g_B)}$,
    \item $\beta_j$ are the coefficients capturing the effect of each feature delta on win probability,
    \item $\varepsilon_m$ is an idiosyncratic error term accounting for unobserved factors influencing match outcomes.
\end{itemize}

\subsection{Residual Calculation}
Following \citet{weidmann2021team}, we derive the Team Player Effect from the residuals of our win prediction model S1, which includes all features capturing task proficiency—specifically, mechanical skill (eAPM), tactical skill (Solo Elo), and functional familiarity. These residuals represent the unexplained variation in match outcomes after accounting for these factors. The residual for team \( g_A \) in match \( m \) is calculated as:

\begin{equation}
r_{mg_A} = y_m - \hat{y}_m
\end{equation}

where:
\begin{itemize}
    \item \( y_m \) is the actual match outcome from the perspective of team \( g_A \) (\( y_m = 1 \) if \( g_A \) wins, \( y_m = 0 \) otherwise),
    \item \( \hat{y}_m \) is the predicted probability of \( y_m = 1 \), obtained from the logistic regression model.
\end{itemize}

Since each individual player participates in multiple matches across different team compositions, their residuals should converge to zero if they solely reflect random noise, given a sufficient number of observations. However, systematic deviations from zero suggest an underlying effect that persists across matches.

These deviations can be interpreted as a player’s consistent contribution to team performance that is not captured by the observed features related to task proficiency, such as mechanical and tactical skill. Following \citet{weidmann2021team}, we define this as the team player effect:

\begin{equation}
    \text{Team Player Effect}_j = \frac{1}{n_j} \sum_{m \in \text{Matches}_j} r_{mg},
\end{equation}

where:
\begin{itemize}
    \item \( n_j \) is the total number of matches player \( j \) has participated in,
    \item \( r_{mg} \) is the residual assigned to the team of player \( j \) in match \( m \).
\end{itemize}

To assess a team's aggregated team player effect, we compute the average team player effect of its members, considering only players who meet a minimum observation threshold \( \tau \). The team player effect for a given team is then defined as:

\begin{equation}
\text{Team Player Effect}_g = \frac{1}{n_g} \sum_{j \in \text{Players}_g} \mathbb{1} \{ n_j \geq \tau \} \cdot \text{Team Player Effect}_j,
\end{equation}

where:
\begin{itemize}
    \item \( n_j \) is the number of matches player \( j \) has participated in,
    \item \( \mathbb{1} \{ n_j \geq \tau \} \) is an indicator function that equals 1 if player \( j \) meets the minimum observation threshold \( \tau \) and 0 otherwise,ensuring that only players with sufficient data contribute to the calculation,
    \item \( n_g = \sum_{j \in \text{Players}_g} \mathbb{1} \{ n_j \geq \tau \} \) represents the number of players in team \( g \) who meet the threshold,
    \item \( \text{Team Player Effect}_j \) represents the individual team player effect of player \( j \).
\end{itemize}

Players with insufficient observations to reliably estimate their team player effect are excluded from the aggregation. The inclusion threshold \( \tau \) is determined based on the distribution of residuals across match counts, using the elbow point of this curve as a cutoff. If no players in a team meet this criterion, a default value of 0 is assigned to the team's team player effect. This approach minimizes noise and ensures that only statistically meaningful residuals contribute to the team-level calculation.

\subsection{Evaluation}

To assess the results, we conduct an analysis of feature contributions, with a particular focus on the Team Player Effect. Additionally, we examine feature interactions to uncover relationships that influence team performance and match outcomes. While our primary analysis relies on joint regressions on the full datasets to ensure representative results, we also perform faceted regression analyses to observe how feature contributions vary under specific contextual conditions, like different team sizes or levels of familiarity. This supplementary approach allows us to explore context-specific variations and compare how feature influences evolve across these facets without compromising the robustness of the main findings derived from the complete dataset.

To ensure the robustness and validity of the residuals, we test for potential biases. Specifically, we evaluate whether the residuals are influenced by contextual factors, such as player positions or team configurations, including cases where teams consist entirely of unfamiliar players. This step is critical to confirm that the Team Player Effect is not systematically biased by specific data characteristics.

To investigate how the general level of team familiarity influences the contribution of the analyzed features, we incorporate interaction terms between absolute team familiarity (measured as the absolute log-transformed value instead of the delta) and the analyzed features team player effect and task proficiency. We apply the same approach for team size and these features, allowing us to analyze whether absolute team familiarity or team size amplify or moderate their influence.

\section{Main Results} \label{sec:results}
In this section, we present the results of our pre-specified models, focusing on the predictive performance of task proficiency, the team player effect, and team familiarity. We first establish the baseline predictions and residual calculations used to define the team player effect. We then examine the final predictive models and assess how team familiarity and team size interact with key features to influence match outcomes.

\subsection{Initial Prediction}

Table \ref{tab:s1_results} reports the results of our initial prediction models (S1.1–S1.3), which serve as the foundation for estimating task proficiency and the team player effect. The models progressively incorporate individual mechanical skill (eAPM), tactical skill (Solo Elo), and functional familiarity features. The coefficients from S1.3 are applied to the same features in data split T2 to construct the task proficiency feature. These estimates form the basis for computing residuals, which capture the unexplained variation in match outcomes and define the team player effect.
\begin{table}[h!]
\smaller
    \caption{Logistic regression results for initial outcome prediction (S1 models). Standard errors (in parentheses) are clustered at the team level. }
    \label{tab:s1_results}
    \begin{minipage}{\columnwidth}
        \begin{center}
            \begin{tabular}{c @{\hskip 12pt} l @{\hskip 15pt} c @{\hskip 15pt} c @{\hskip 15pt} c}
                \toprule
                & \textbf{Feature} & \textbf{S1.1} & \textbf{S1.2} & \textbf{S1.3} \\
                \midrule
                \multirow{2}{*}{\centering \textit{Solo Features}} 
                & eAPM & \textbf{0.2942}*** (0.0024) & \textbf{0.1618}*** (0.0025) & \textbf{0.1516}*** (0.0025) \\
                & Solo Elo & — & \textbf{0.6216}*** (0.0029) & \textbf{0.6133}*** (0.0030) \\
                \hdashline
                \multirow{3}{*}{\centering \textit{Functional Familiarity}} 
                & Match Count & — & — & \textbf{-0.0506}*** (0.0033) \\
                & Map & — & — & \textbf{0.1588}*** (0.0036) \\
                & Civilization & — & — & \textbf{0.0383}*** (0.0034) \\
                \hdashline
                & Constant & \textbf{0.0007} (0.0024) & \textbf{-0.0002} (0.0025) & \textbf{-0.0000} (0.0025) \\
                \midrule
                & Observations & 811,914 & 811,914 & 811,914 \\
                & \textit{Pseudo $R^2$} & 0.0161 & 0.0700 & 0.0744 \\
                \bottomrule
            \end{tabular}
            \\
                    \vspace{5pt}
    \footnotesize \emph{Note:} Significance levels: * $p < 0.05$, ** $p < 0.01$, *** $p < 0.001$.
        \end{center}
    \end{minipage}

\end{table}

\subsection{Residual Calculation}

\begin{figure}[htbp]
    \centering
    \begin{minipage}{0.48\textwidth}
        \centering
        \includegraphics[width=\linewidth]{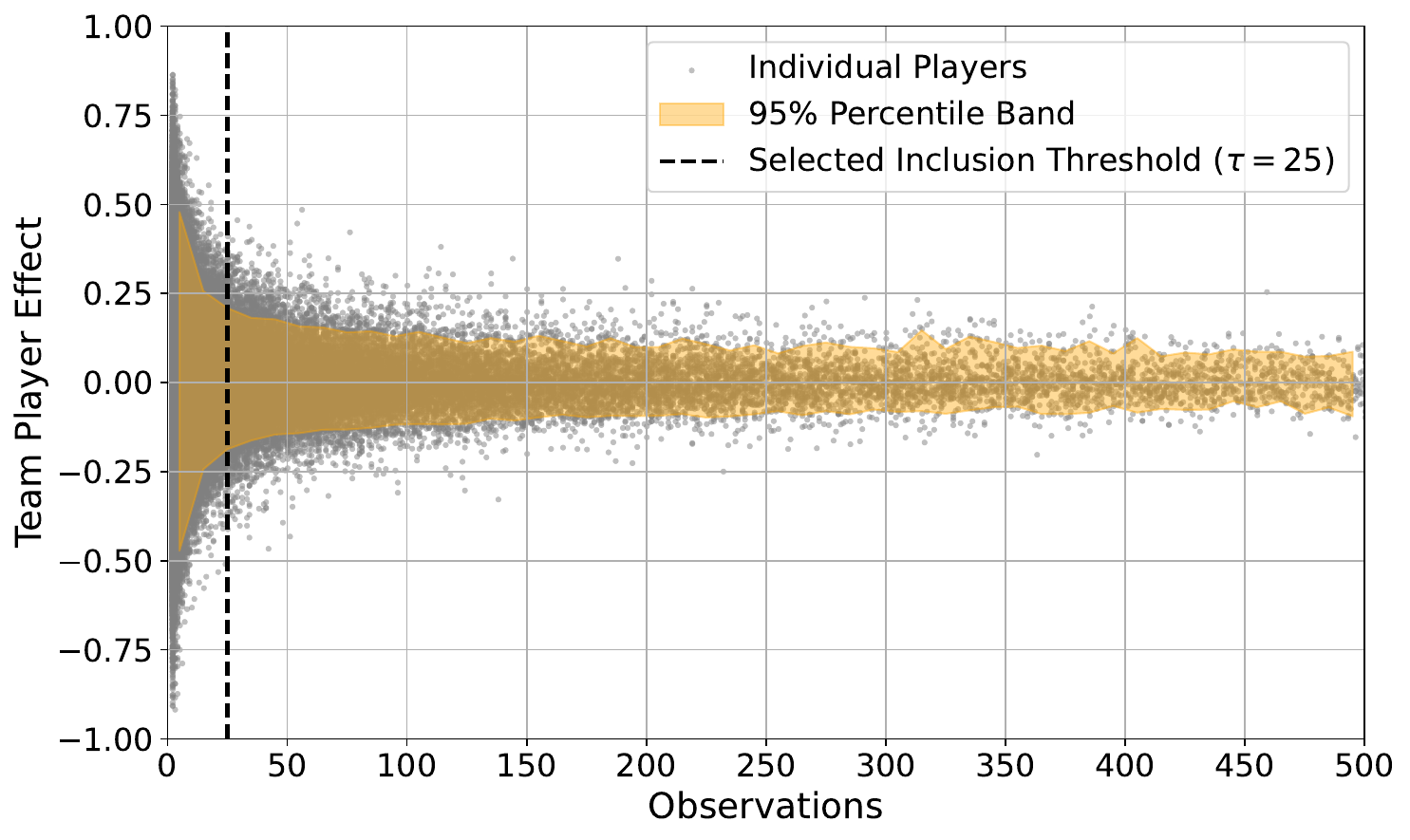}
        \caption{Residuals of the team player effect over the number of observations used for its estimation with a 95\% percentile band (binsize=10).}
        \label{fig:residual_vs_match_count}
    \end{minipage}
    \hfill
    \begin{minipage}{0.48\textwidth}
        \centering
        \includegraphics[width=\linewidth]{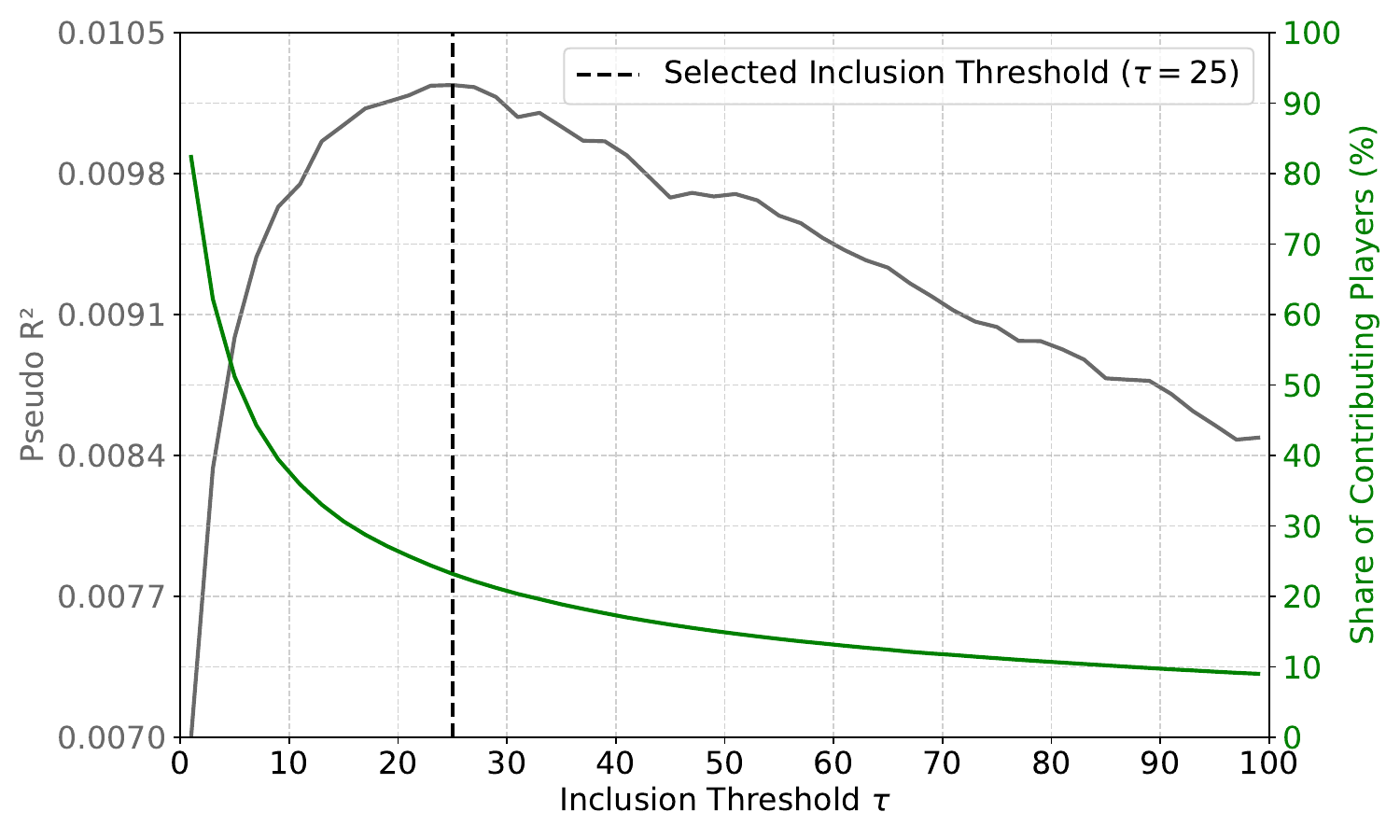}
        \caption{Explanatory performance of models based solely on the team player effect across different inclusion thresholds \( \tau \) (stepsize=2).}
        \label{fig:inclusion_threshold}
    \end{minipage}
\end{figure}

Figure \ref{fig:residual_vs_match_count} presents the residuals of the team player effect for all players in Split T1 against the number of observations used for estimation on an exemplary fold. These residuals are derived from the difference between the actual outcomes and the predictions of S1.3. Each point represents an individual player, and the shaded band denotes the 95\% percentile range of residuals within bins of 10 observations. As the number of matches increases, the residuals stabilize within a narrower range, indicating improved estimation reliability. The inclusion threshold \( \tau = 25 \), highlighted as a vertical dashed line, is selected based on the elbow point of this convergence pattern. This threshold balances statistical robustness and sample coverage by filtering out residuals with high variability due to insufficient data. Figure \ref{fig:inclusion_threshold} validates this choice by assessing the explanatory power of a logistic win prediction model on Split T2, using only the team player effect as a feature. The x-axis represents different inclusion thresholds \( \tau \) applied in the construction of the team player effect, while the model’s explanatory power peaks at the selected threshold. The right y-axis captures the share of players in the sample who meet the threshold, showing a steady decline as \( \tau \) increases. This ensures maximal predictive strength while retaining information on 23\% of the 116,833 players in the sample. The share does not start at 100\% because only 82\% of players in Split T2 are also featured in Split T1, preventing team player effect computation even at the lowest threshold.

To ensure robustness we also calculate residuals separately for players in the pocket and flank positions. The residuals for these two positions exhibit a high degree of correlation, which increases with the number of observations. As the sample size grows, the residuals converge, indicating no systematic differences in structure between preferred player roles. To further evaluate the residual properties, we assess whether the distribution of S1.3 residuals differs systematically between matches where all players have zero prior familiarity and the full dataset. Given that match assignments are randomized within the ranked matchmaking system, any systematic deviation in these residuals could indicate a structural bias in the estimation. We apply Kolmogorov-Smirnov tests across various match count bins to compare the distributions of residuals in both subsets. The results show no significant differences between the residual distributions of zero-familiarity teams and the overall match set, even as sample sizes vary. This suggests that the residuals are robust and remain unaffected by the absence of familiarity in team compositions, further validating the approach.

\subsection{Final Prediction}

\begin{figure}[!h]
\centering
	\includegraphics[width=.5\textwidth]{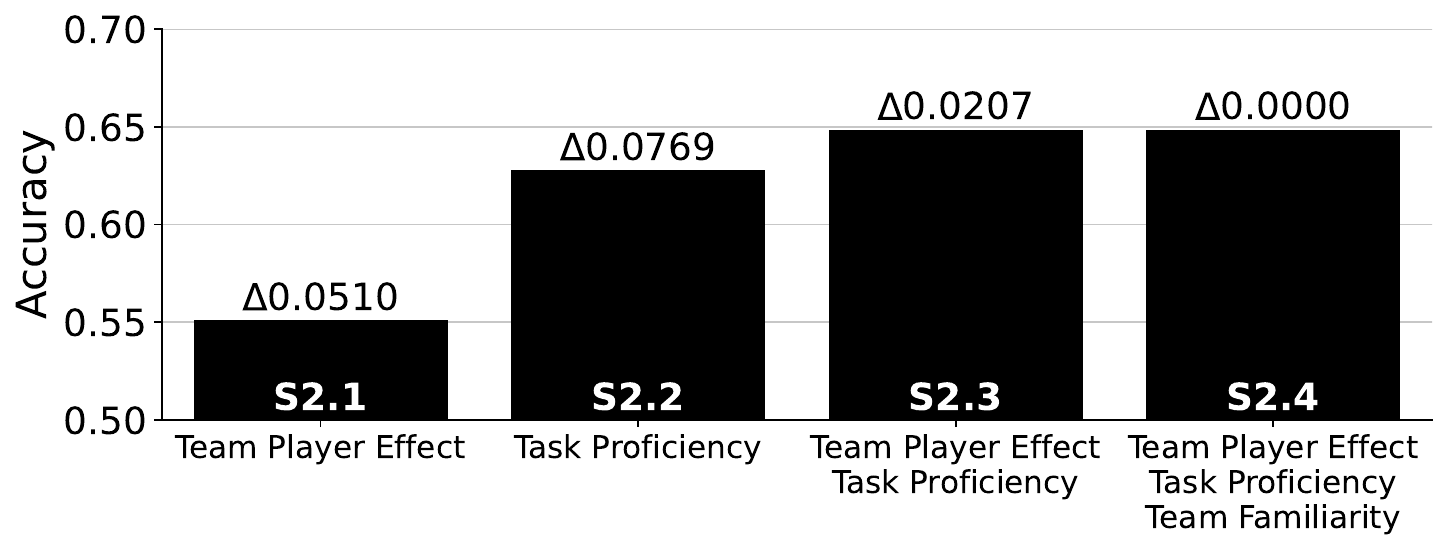}
        \caption{Model accuracy on the test set of Split T2. $\Delta$ values indicate accuracy improvements over the previous model, with S2.1 compared to a random guess.}
	\label{fig:accuracy_plot}
\end{figure}

The final predictions S2 were conducted on Split T2 using the feature sets S2.1 through S2.4, the results are reported in Table \ref{tab:s2_results}. The peformance on the test set is depicted in Figure \ref{fig:accuracy_plot}. The findings demonstrate a consistent improvement in predictive performance as additional features are incorporated. Each added feature contributes significant explanatory power, with the exception of team familiarity.

Although the team familiarity delta lacks explanatory power for outcome prediction, the absolute level of team familiarity influences performance dynamics through its interactions with key features. To quantify these effects, we compute Marginal Effects at the Mean (MEM) using a logistic regression model. MEM estimates the partial derivatives of the probability of winning with respect to each interaction term while holding all other variables at their mean values. This approach provides an interpretable measure of how each interaction affects the likelihood of winning under average conditions. Standard errors and significance levels are derived using robust estimation methods.

\begin{table}[h]
\smaller
    \caption{Logistic regression results for final outcome prediction (S2 models). Standard errors (in parentheses) are clustered at the team level.}
    \label{tab:s2_results}
    \begin{minipage}{\columnwidth}
        \begin{center}
            \begin{tabular}{l @{\hskip 15pt} c @{\hskip 15pt} c @{\hskip 15pt} c @{\hskip 15pt} c}
                \toprule
                \textbf{Feature} & \textbf{S2.1} & \textbf{S2.2} & \textbf{S2.3} & \textbf{S2.4} \\
                \midrule
                \textbf{Team Player Effect} & \textbf{0.2454}*** & — & \textbf{0.4256}*** & \textbf{0.4265}*** \\
                & (0.0028) & — & (0.0030) & (0.0030) \\
                \textbf{Task Proficiency} & — & \textbf{0.7157}*** & \textbf{0.8300}*** & \textbf{0.8266}*** \\
                & — & (0.0032) & (0.0034) & (0.0034) \\
                \textbf{Team Familiarity} & — & — & — & \textbf{0.0142}*** \\
                & — & — & — & (0.0027) \\
                Constant & \textbf{-0.0001} & \textbf{-0.0000} & \textbf{-0.0002} & \textbf{-0.0002} \\
                & (0.0027) & (0.0028) & (0.0027) & (0.0027) \\
                \midrule
                Observations & 710,817 & 710,817 & 710,817 & 710,817 \\
                \textit{Pseudo $R^2$} & 0.0106 & 0.0743 & 0.1003 & 0.1004 \\
                \bottomrule
            \end{tabular}
            \\
        \vspace{5pt}
        \footnotesize \textit{Note:} Significance levels: * $p < 0.05$, ** $p < 0.01$, *** $p < 0.001$.
        \end{center}

    \end{minipage}

\end{table}

Table \ref{tab:marginal_effects_interactions} presents these results, showing that absolute team familiarity significantly interacts with both the team player effect and task proficiency. The interaction between absolute team familiarity and the team player effect is both statistically significant (0.0155) and notable relative to the baseline coefficient of the team player effect in S2.4 (0.4265), suggesting that familiarity amplifies its impact on match outcomes. Similarly, the interaction with task proficiency is positive and slightly higher in magnitude (0.0212), aligning with task proficiency’s overall stronger predictive role in match outcomes.  Together, these results indicate that familiarity enhances both social and technical aspects of team coordination, shaping the way individual skills and team dynamics contribute to performance. A similar pattern emerges for team size. To capture this relationship, team size is treated as a continuous feature, and its interactions with the key predictors are analyzed. The interaction between team size and the team player effect is positive (0.0353), reinforcing the idea that the impact of unobserved team-related contributions strengthens as teams grow larger. 	Conversely, team size negatively interacts with task proficiency (-0.0494), suggesting that the influence of individual task proficiency on match outcomes diminishes in relative importance as team size increases, reflecting a shift toward collective coordination.

\begin{table}[h]
\smaller
    \caption{Marginal Effects at Mean (MEM) for a Model Including Feature Set of S2.4 with Interaction Terms. Standard errors (in parentheses) are clustered at the team level.}
    \label{tab:marginal_effects_interactions}
    \begin{minipage}{\columnwidth}
        \begin{center}
            \begin{tabular}{l @{\hskip 15pt} c @{\hskip 15pt} c}
                \toprule
                \textbf{Feature / Interaction Term} & \textbf{Marginal Effect} & \textbf{(Std. Error)} \\
                \midrule
                \textbf{Main Features} \\
                Task Proficiency & \textbf{0.2530}*** & (0.0034) \\
                Team Player Effect & \textbf{0.0767}*** & (0.0029) \\
                Team Familiarity & \textbf{0.0055}*** & (0.0007) \\
                \midrule
                \textbf{Abs. Team Familiarity Interactions} \\
                Abs. Team Familiarity × Team Player Effect & \textbf{0.0155}*** & (0.0008) \\
                Abs. Team Familiarity × Task Proficiency & \textbf{0.0212}*** & (0.0009) \\
                \midrule
                \textbf{Team Size Interactions} \\
                Team Size × Team Player Effect  & \textbf{0.0353}*** & (0.0030) \\
                Team Size × Task Proficiency & \textbf{-0.0494}*** & (0.0034) \\
                \bottomrule
            \end{tabular}
        \\
        \vspace{5pt}
        \footnotesize \textit{Note:} Significance levels: * $p < 0.05$, ** $p < 0.01$, *** $p < 0.001$.
        \end{center}
    \end{minipage}

\end{table}

\begin{figure}[h]
    \centering
    \includegraphics[width=\textwidth]{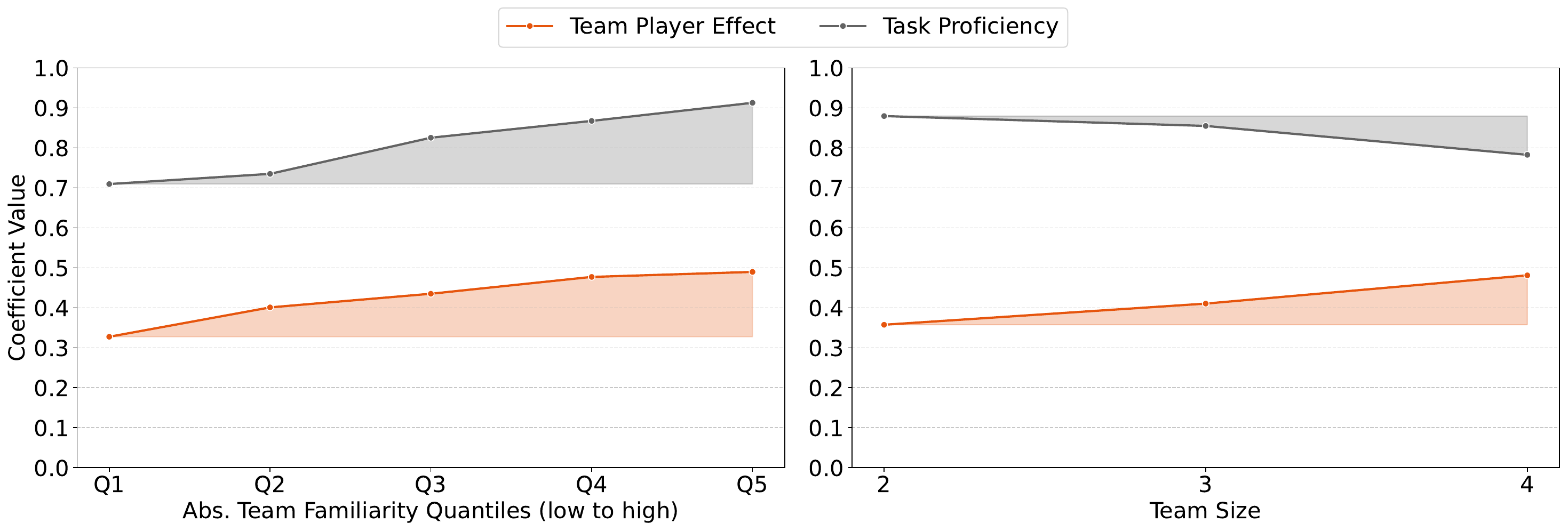}
    \caption{S.24 Feature Coefficients across Familiarity Quantiles and Team Size. 
    (Left) Coefficient trends across familiarity quantiles, illustrating the impact of increasing familiarity levels on feature importance. 
    (Right) Coefficient trends by team size, showing how feature contributions change as the team size increases. Shaded areas represent the magnitude of variation.}
    \label{fig:coeff_dev}
\end{figure}

To further illustrate these findings, Figure \ref{fig:coeff_dev} examines how the coefficients of task proficiency and the team player effect in S2.4 vary across different levels of familiarity and team sizes. The left panel divides the dataset into five quantiles based on increasing levels of absolute team familiarity and estimates independent regressions for each quantile. This analysis reveals a clear pattern in which the team player effect gains importance as familiarity rises, with its relative increase exceeding that of task proficiency. The right panel provides a complementary perspective by splitting the dataset based on team size rather than familiarity levels. Here, task proficiency remains the dominant predictor but gradually declines in influence as team size increases, while the team player effect strengthens. 

\section{Discussion} \label{sec:discussion}

Our findings provide empirical evidence for the existence of a team player effect—unobserved but systematic component of team performance beyond individual task proficiency. We find that some individuals consistently enhance team performance beyond what is expected based on their task proficiency alone, an effect that remains stable across different settings and explains a substantial share of team success. By employing a residual-based estimation approach, we isolate this effect, and robustness checks confirm its persistence. A one-standard-deviation increase in the team player effect is associated with a 0.43 increase in the log-odds of winning, corresponding to a 54\% increase in the odds of winning (Table \ref{tab:s2_results}, S2.4). Figure \ref{fig:residual_vs_match_count} shows that deviations from zero remain stable even at high observation counts, suggesting that the team player effect reflects a persistent individual trait rather than statistical noise. The explanatory power of models incorporating this measure further validates its predictive relevance, particularly after accounting for task proficiency.

A key insight from our analysis is that a minimum number of observations is required for a reliable estimation of the team player effect. Figure \ref{fig:inclusion_threshold} demonstrates that explanatory power peaks at the selected threshold, balancing statistical precision with sample representativeness. This ensures that residual-based team player estimates capture systematic contributions rather than idiosyncratic fluctuations. Moreover, robustness tests confirm that these residuals do not systematically differ between zero-familiarity teams and the full dataset, suggesting that the estimation method is not biased by team composition effects. The high correlation of residuals across in-game roles further supports the generalizability of the team player effect across different strategic positions.

Beyond establishing the team player effect, our results highlight the role of team familiarity in shaping the impact of individual attributes on performance. While familiarity does not directly improve predictive accuracy (Figure \ref{fig:accuracy_plot}), teams with prior shared experience benefit more from the presence of team players. This finding aligns with the view that social skills—encompassing both observable social intelligence and the latent team player effect—facilitate structured coordination and role alignment. As \citet{weidmann2021team} suggest, social skills enhance team performance as much as IQ, and our results indicate that familiarity amplifies their effectiveness. Specifically, repeated interactions strengthen both social and technical contributions, reinforcing the idea that familiarity fosters structured coordination and efficient role allocation \citep{ching2021extemporaneous,ching2024competitive}.

These results suggest that task proficiency and the team player effect serve as channels through which underlying individual skills influence match outcomes, with familiarity moderating their effectiveness. Rather than directly determining success, familiarity strengthens the complementarity between individual contributions, improving coordination and reducing inefficiencies in team production. This aligns with broader theories of team production, where repeated interactions foster implicit knowledge and increase the returns to social capital and cooperative specialization \citep{massaro2020intellectual,mach2010differential}.

The interaction between the team player effect and familiarity further raises the possibility that reputation mechanisms contribute to team performance. Familiarity may serve as a proxy for implicit trust among teammates, allowing players to rely more on unobserved social contributions. This interpretation is consistent with prior work suggesting that familiarity fosters cooperation and more effective role allocation within teams \citep{mach2010differential,akcsin2021learning,maynard2019really}. Future research could further investigate whether familiarity improves coordination through repeated exposure or whether it reflects deeper relational bonds between teammates.

Finally, our results show that as team size increases, social skills play an increasingly important role in shaping team performance. Figure \ref{fig:coeff_dev} shows that as team size increases, the relative importance of task proficiency declines, while the contribution of the team player effect strengthens. This suggests that in larger teams, coordination and adaptability become increasingly important, while individual execution plays a relatively diminished role. Table \ref{tab:marginal_effects_interactions} further supports this finding: the positive interaction between team size and the team player effect (0.0353) indicates that the value of unobserved team-related contributions becomes more pronounced as teams grow. Conversely, the negative interaction between team size and task proficiency (-0.0494) suggests that individual skill exerts less influence on outcomes in larger teams. These patterns are consistent with theories of coordination costs and diminishing marginal returns to individual skill in larger groups. As teams expand, the ability to function within a collective structure gains relevance, reinforcing the shift from individually driven performance toward a more cohesive mode of play.







\section{Conclusion and Outlook} \label{sec:conclusion}
We provide large-scale empirical evidence on the role of social skills and familiarity in team-based production, expanding on prior research that identifies team players—individuals who systematically enhance team performance beyond their task proficiency alone \citep{weidmann2021team}. We confirm the existence of the team player effect in a high-stakes, dynamically evolving environment, moving beyond controlled laboratory experiments to a large-scale, naturalistic setting. In line with prior research, we find that some individuals consistently improve team outcomes, even when controlling for mechanical skill, tactical ability, and contextual familiarity. The persistence of this effect across millions of experiments where team members were exogenously assigned to teams under quasi-random matchmaking highlights the robustness of social skills as a key driver of coordination and productivity.

Beyond confirming the presence of the team player effect, our findings highlight the complementarity between social skills and team familiarity. While prior research has examined these factors separately—social skills through the team player effect \citep{weidmann2021team} and team familiarity in relation to specialization \citep{ching2021extemporaneous}—our results show that they interact to enhance team performance. Specifically, teams with prior shared experience benefit more from the presence of team players, suggesting that familiarity amplifies the effectiveness of social skills by facilitating coordination and role alignment. This finding implies that prior interactions enable socially skilled individuals to contribute more effectively, strengthening team cohesion and performance. Additionally, we show that the role of social skills becomes more pronounced as team size increases. As teams and complexity grow, individual execution plays a relatively smaller role, while structured coordination gains importance \citep{akcsin2021learning,neuman1999team,ahearn2004leader}. Our results indicate that socially skilled individuals facilitate this process, helping teams navigate coordination frictions more effectively. While familiarity strengthens the effectiveness of social skills by fostering structured coordination, team size introduces an additional dimension to this relationship. As teams grow larger, coordination challenges intensify, shifting the balance from individual execution toward structured collaboration \citep{mao2016experimental,lepine2008meta,yuan2022leader}. Our results indicate that socially skilled individuals play a crucial role in overcoming these frictions, reinforcing their growing importance in larger teams.

These findings contribute to the broader literature on teamwork, social skills, and familiarity in economic settings while offering practical implications for organizational design and human capital management. First, we extend prior research on the team player effect by demonstrating its robustness in a large-scale, high-pressure environment. Unlike studies relying on controlled laboratory experiments, our findings leverage millions of quasi-randomly assigned team interactions, offering a naturalistic test of social skills in team coordination. Second, we show that the benefits of social skills are amplified by familiarity, reinforcing the role of prior shared experience in facilitating structured coordination. Third, we find that the importance of social skills increases with team size, suggesting that coordination challenges become more pronounced in larger teams, where individual execution is less decisive.

While teams in online gaming environments are inherently interesting, the findings of this research are expected to generalize to other contexts, such as non-hierarchical teams in organizations and online labor platforms. The ability to identify and recruit team players presents a significant opportunity for firms seeking to improve team-based productivity. Our results suggest that large productivity gains are possible for employers who can systematically identify individuals with strong social skills, particularly in contexts where teams are frequently reconfigured. Furthermore, we provide empirical guidance on the feasibility of such identification: in our setting, a minimum of 25 interactions is required to reliably estimate a player’s team contributions, offering a benchmark for assessing social skills in other team-based environments. Finally, our results highlight the importance of team familiarity and team size in shaping the impact of social skills, underscoring the role of structured coordination in driving team performance.

Beyond these immediate contributions, our study highlights opportunities for future research. Given that our dataset includes persistent player identifiers and feasible means of contact, future research could directly engage with identified team players to assess their social skills, cognitive attributes, and professional backgrounds, as well as their roles within and the characteristics of the teams they participate in. Understanding how these individuals contribute to team dynamics across different contexts could provide further insights into their long-term effectiveness and implications for hiring and team formation. For instance, administering social intelligence tests or surveying professional occupations could offer richer insights into how team player traits translate across different domains.




\bibliographystyle{unsrtnat}
\bibliography{references}

\clearpage
\clearpage

\appendix

\section{Technical Appendix}

\begin{figure}[h]
    \centering
    \begin{minipage}{0.48\textwidth}
        \includegraphics[width=\textwidth]{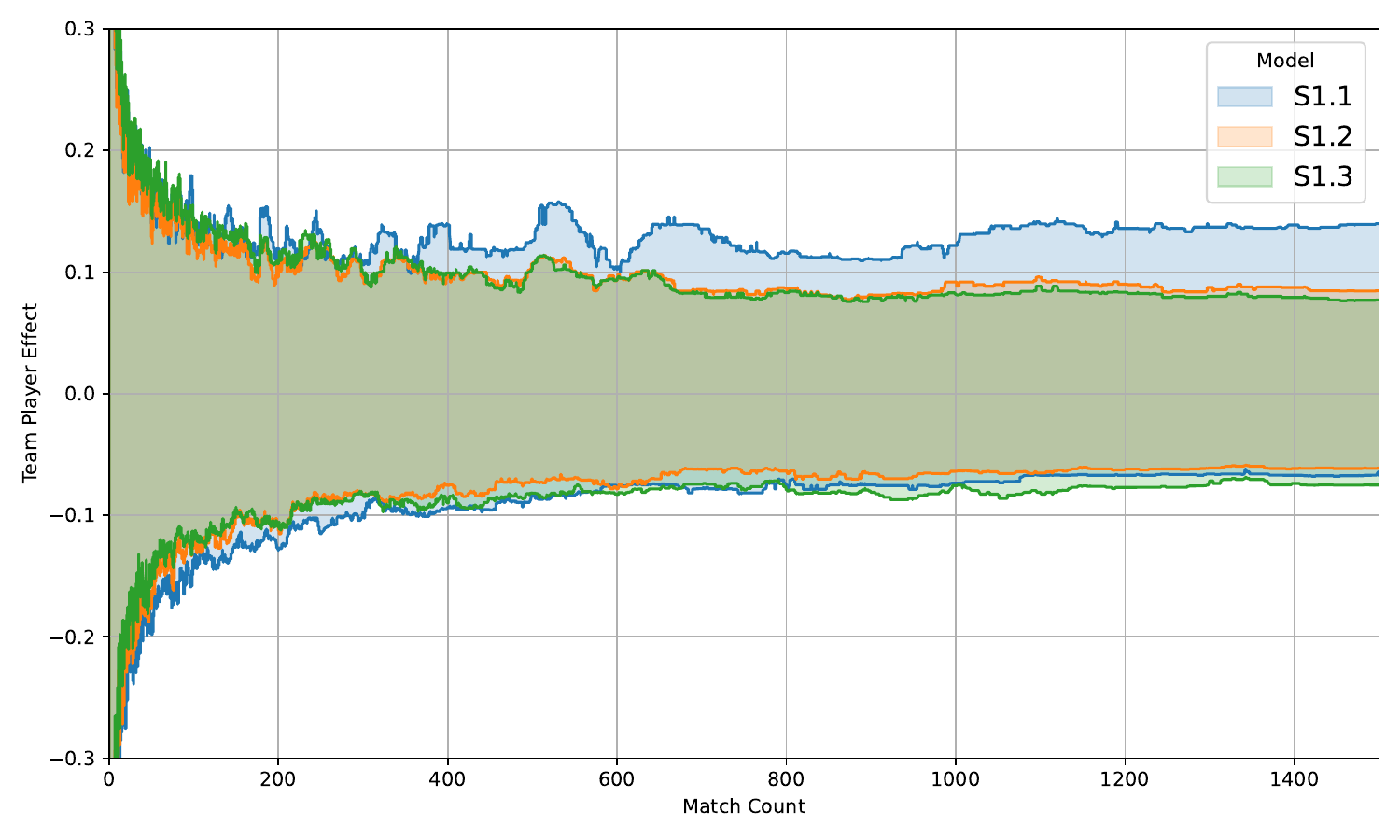}
    \end{minipage}
    \hfill
    \begin{minipage}{0.48\textwidth}
        \includegraphics[width=\textwidth]{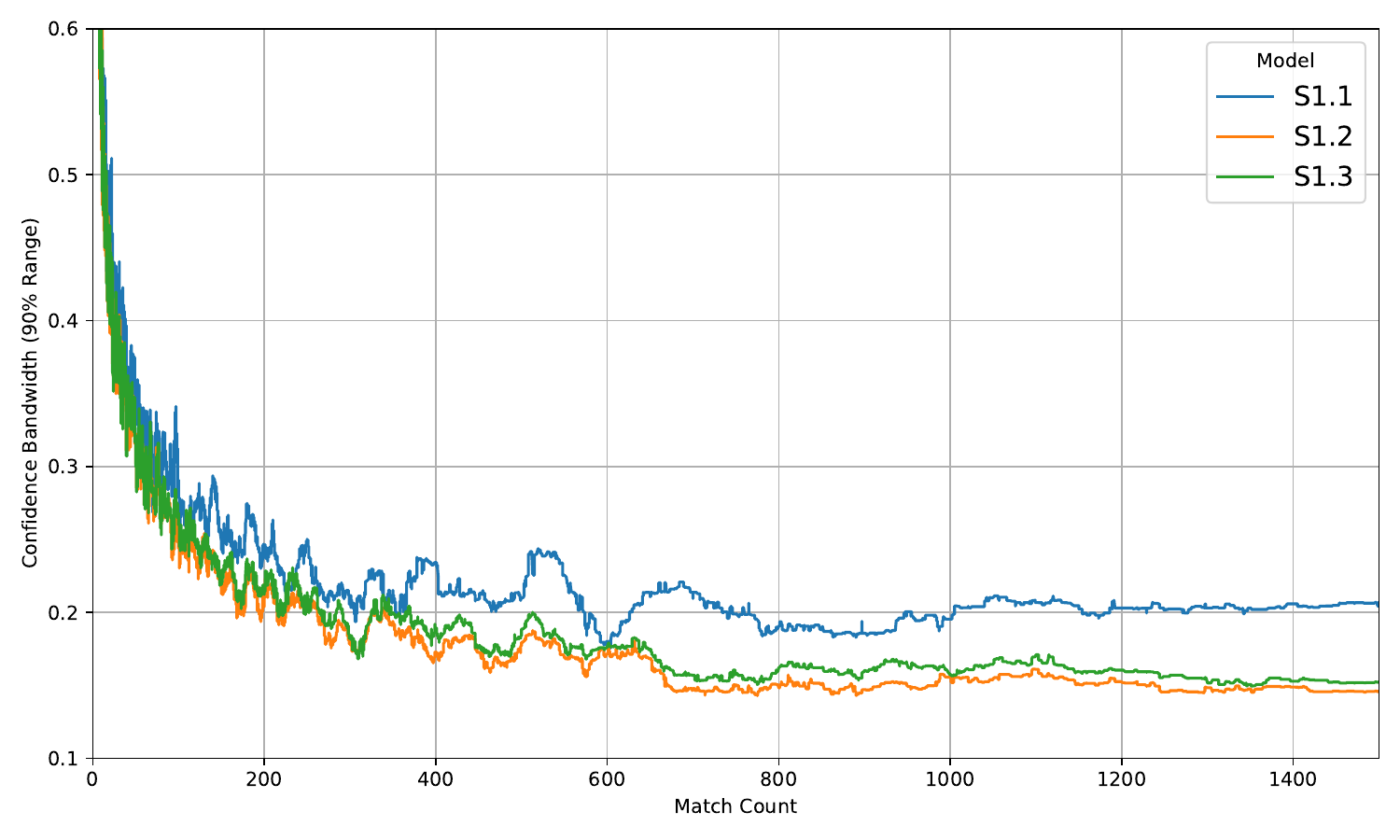}
    \end{minipage}
	\caption{(Left) 95th percentile of residuals for S1.1-S1.3 over match count; (Right) residual bandwidth 95th for S1.1-S1.3, both computed using a rolling window of 500 observations.}

    \label{fig:m1-m3-res}
\end{figure}

This appendix examines residual variability across models S1.1–S1.3, highlighting how different feature sets impact explanatory power. Figure \ref{fig:m1-m3-res} illustrates two aspects of residual behavior: the left panel shows the 95th percentile of residuals as a function of match count, while the right panel tracks the residual bandwidth (95\% range) across models.

As expected, residual variability decreases with more observations, confirming the robustness of our estimation approach. The largest reduction in residuals occurs between S1.1 and S1.2, driven by the inclusion of Solo Elo, which captures individual tactical skill. However, the transition from S1.2 to S1.3 does not further reduce residuals, indicating that the added functional familiarity features—such as experience with specific civilizations and maps—do not contribute to systematic individual differences across contexts. Instead, they primarily reflect situational expertise, aligning with our interpretation that S1.3 fully captures task proficiency but does not diminish unexplained individual contributions.

Notably, even at higher match counts, residuals do not converge to zero. This persistence supports our identification of the team player effect as an individual-specific factor not accounted for by task proficiency alone, reinforcing its role as an unobserved but systematic determinant of team success.

\end{document}